\documentclass[epj]{svjour}
\usepackage{graphicx}
\usepackage{amsmath}
\usepackage{amssymb}
\usepackage{mathtext}

\begin{document}

\onecolumn

\title{Effect of Temperature Wave on Diffusive Transport of Weakly-Soluble
       Substances in Liquid-Saturated Porous Media}
\titlerunning{Temperature Wave \& Diffusion in Saturated
       Solutions Filling Porous Media}

\author{Pavel~V.~Krauzin\inst{1} \and
 Denis~S.~Goldobin\inst{1,2,3}}

\authorrunning{P.~V.~Krauzin, D.~S.~Goldobin}

\institute{Institute of Continuous Media Mechanics, UB RAS,
 Perm 614013, Russia
\and Department of Mathematics, University of Leicester,
 Leicester LE1 7RH, UK
\and Department of Theoretical Physics, Perm State University,
 Perm 614990, Russia}

\date{\today}

\abstract{
We study the effect of surface temperature oscillations on
diffusive transport of solutes of weakly-soluble substances
through liquid-saturated porous media. Temperature wave induced by
these oscillations and decaying deep in the porous massif creates
the solubility wave along with the corresponding solute diffusion
flux wave. When the non-dissolved fraction is immobilized in
pores---for gases the bubbles can be immobilized by the surface
tension force, for solids (e.g., limestone, gas-hydrates) the
immobilization of non-dissolved phase is obvious---the only
remaining mechanisms of mass transport are related to solute flux
through liquid in pores. We evaluate analytically the generated
time-average mass flux for the case of medium everywhere littered
with non-dissolved phase and reveal the significant effect of the
temperature wave on the substance release from the massif and
non-dissolved mass redistribution within the massif. Analytical
theory is validated with numerical calculations.
}

\maketitle

\section{Introduction}
Transport of gases, as well as of any weakly-soluble substances,
in liquid-saturated porous media appears to possess unique
features~\cite{Donaldson-etal-1997,Donaldson-etal-1998,Haacke-Westbrook-Riley-2008,Goldobin-Brilliantov-2011}.
Specifically, for a pore diameter small enough the pore-sized
bubbles are immobilized by the surface tension force. The critical
pore size can be readily accessed from the balance of the surface
tension and buoyancy forces, which yields
the value of order of $1\,\mathrm{mm}$ for the air-water system.
Meanwhile, the bubbles which are large compared to the pore size
will experience splitting during displacement due to the fingering
instability~\cite{Lyubimov-etal-2009}. As a result, the
hydrodynamic transport of the gas phase through porous media
becomes practically impossible for a small volumetric fraction of
gas in pores ({\it or} gas saturation). The critical
immobilisation gas saturation varies from system to system and
depends on the time scale (e.g., the leakage process significant
on the time scales of millions of years is negligible for
processes evolving in days), but remains within the range from
0.5--1\%~\cite{Firoozabadi-Ottesen-Mikkelsen-1992} to several
percent (see, e.g.,~\cite{Moulu-1989}). The critical immobilized
gas mass is at least one order of magnitude larger than the
feasible variations of gas mass dissolved in a saturated aqueous
solution of oxygen, nitrogen, or methane under the Earth's
near-surface conditions for pressure and temperature. The latter
conditions are relevant for the processes in wetlands and
peatbogs. Thus the principal mechanism of the gas mass transport
in groundwater turns out to be the transport of gas molecules
dissolved in water.

In the massif zone where the aqueous solution is saturated---i.e.,
porous massif is littered with bubbles of gas phase---the gas flux
is solely determined by the field of solubility, which depends on
pressure and
temperature~\cite{Pierotti-1976,Goldobin-Brilliantov-2011}. In
this relation it is remarkable that the temperature increase from
$0^\circ\mathrm{C}$ to $20^\circ\mathrm{C}$ leads to the decrease
of the solubility of the main atmospheric gases and methane by a
factor $1.5$. This means that the annual temperature wave
propagating into the porous massif (e.g., see~\cite{Yershov-1998})
gives rise to the solubility wave of significant amplitude and
associated diffusive fluxes. Hence, the annual temperature wave
can significantly affect the processes of saturating groundwater
with gases or methane release from wetlands and peatbogs.

The phenomenon under consideration is common and can be
significant for various systems with different origins of the
surface temperature oscillations, including technological systems
(filters, porous bodies of nuclear and chemical reactors, etc.).
However, for the sake of convenience, in this paper, we first
focus on the case featured by the hydrostatic pressure gradient
which is significant for geological systems, where pressure
doubles on the depth of 10 meters, leading to significant change
of solubility. The no pressure gradient case can be derived from
the case of hydrostatic pressure by setting the gravity $g$ in
resulting analytical equations to zero.

Our consideration will not be restricted to gases only.
Weakly-soluble solids are very common in natural systems; these
are limestone and other weakly-soluble mineral salts, methane
hydrate and hydrates of other gases, etc. For these substances the
immobilization of non-dissolved phase in pores is obvious. The
amount of matter which can be held in the solution in pores
compared to the amount of matter which can stay in the
non-dissolved phase in the same pores is as small as for gases or
even smaller. Solubility of these substances depends on
temperature as strongly as for gases. In contrast to gases,
however, the solubility of solids is not proportional to pressure
but nearly independent of it. Nonetheless, the pressure
inhomogeneity is owned by the hydrostatic pressure gradient, and
the analytical results we will derive for gases can be downgraded
to the case of solids by setting the gravity $g$ to zero, which
will eliminate the pressure inhomogeneity effect on solubility,
and proper correction of the temperature dependence of solubility.
Thus from the view point of physics the theory for weakly-soluble
solids turns out to be a specific case of the theory for gases.

We restrict our consideration to the case when the solution is
saturated everywhere in the massif. This restriction is suggested
not only by the motive of fundamental theoretical interest related
to problem novelty but also by its practical relevance. It is
relevant for the methane release from peatbogs, as the process of
methane generation there can be intense enough to maintain the
saturation~\cite{methane-peatbogs-1,methane-peatbogs-2}, and for
atmospheric gases. For the latter, it is important that the
present time is the warm epoch against the background of the
``main'' cold state of 100\,000-year Glacial-Interglacial
cycles~\cite{iceage-1,iceage-2}. During the cold period the
groundwater is saturated with atmospheric gases under enhanced
solubility conditions (low temperature), while during the warm
period of diminished solubility the excessive gas mass from the
solution forms gaseous bubbles in pores. For low-permeability
fractureless massifs, the dominant transport mechanism is the
molecular diffusion. Molecular diffusion operates with equal
efficiency downwards, during the cold periods when the groundwater
is being saturated with gases, and upwards, during the warm
periods when the system is losing the gas mass. The longer the
saturating period is, the longer duration of the release period is
needed to remove the excessive amount of gases. Due to asymmetry
between cold (long) and warm (short) periods groundwater in the
near-surface layers should remain gas-saturated during the warm
period. Thus, for the current time, one should observe gradual
release of the previously saturated gases from massifs under the
saturation condition. Short-time temperature waves can affect the
rate of this release. Thus, we can see the relevance of the
problem statement assuming the solution to be everywhere
saturated.

With weakly-soluble solids, the effect can be potentially utilized
in technology for filling the porous matrix with some
weakly-soluble ``guest'' substance; the spatial mass distribution
pattern of the ``guest'' substance can be controlled by the
surface temperature waveform. For natural methane hydrate deposits
in seabed sediments, the effect of temperature waves on the
deposit and the gas release from it is of interest in relation to
the natural Glacial-Interglacial temperature cycles and potential
global climate change~\cite{Hunter-etal-2013}.

In this paper, we calculate the diffusion flux of a weakly-soluble
substance in the presence of temperature wave and redistribution
of the mass in the non-dissolved phase. The consideration is
firstly performed for gases and then downgraded to the case of
solids. Both the cases of molecular diffusion and hydrodynamic
dispersion are considered. The analytical results are validated
with numerical simulation. The periods of negative temperatures
with frozen groundwater are beyond the scope of this study and
will be considered elsewhere.

\section{Temperature wave in porous media with non-dissolved phase}
\begin{figure}[!t]
\center{
\includegraphics[width=0.35\textwidth]%
 {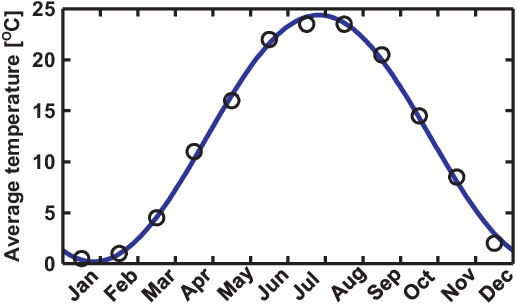}
}

  \caption{(Color online)
Average temperature (corresponds to the surface soil temperature
$T(z=0)$, eq.\,(\ref{eq01})) in New York City (circles) and its
harmonic approximation (solid line) [data from open sources]
 }
  \label{fig1}
\end{figure}

We adopt harmonic annual oscillation of the surface temperature,
$T_0+\Theta_0\cos{\omega t}$, where $T_0$ is the annual-mean
temperature, $\Theta_0$ is the oscillation amplitude, and
$\omega=2\pi/\mathrm{year}$. This assumption is accurate enough
due to two reasons: (i) real temperature records slightly deviate
from their harmonic reduction (e.g., see~\cite{Yershov-1998} and
fig.\,\ref{fig1}) and (ii) the penetration depth of higher
harmonics into massif rapidly decreases (see eq.\,(\ref{eq01})
below). With the specified surface temperature the temperature
field in the half-space, which is governed by the heat diffusion
equation
 $\partial T/\partial t=\chi\Delta T$
with no heat flux condition at infinity, is
\begin{equation}
T(z)=T_0+\Theta_0e^{-kz}\cos(\omega t-kz)\,,
\quad
k=\sqrt{\omega/2\chi}\,.
\label{eq01}
\end{equation}
Here $\chi$ is the heat diffusivity, the $z$-axis is oriented
downwards and its origin is on the massif surface. The pressure
field is hydrostatic
\begin{equation}
P=P_0+\rho gz\,,
\label{eq02}
\end{equation}
where $P_0$ is the atmospheric pressure, $\rho$ is the liquid
(mostly, water) density, and $g$ is the gravity.

For pressure up to $10\,\mathrm{atm}$ and far from the solvent
boiling temperature the solubility depends on temperature $T$ and
pressure $P$ as follows~\cite{Pierotti-1976}
\begin{equation}
 X^{(0)}(T,P) \simeq
 X^{(0)}(T_0,P_0)\frac{T_0}{T}\frac{P}{P_0}
 \exp\left[q\left(\frac{1}{T}-\frac{1}{T_0}\right)\right]\,,
\label{eq03}
\end{equation}
where molar solubility $X^{(0)}$ is the molar amount of solute per
1 mole of solvent, $T_0$ and $P_0$ are reference values, the
choice of which is guided merely by convenience reason, and
$X^{(0)}(T_0,P_0)$ is the solubility at the reference temperature
and pressure; the parameter $q\equiv-G_i/k_\mathrm{B}$, with $G_i$
being the interaction energy between a solute molecule and the
surrounding solvent molecules, is provided in table~\ref{params}.
For solids the solubility approximately reads
\begin{equation}
 X_\mathrm{solid}^{(0)}(T,P) \simeq
 X_\mathrm{solid}^{(0)}(T_0,P_0)
 \exp\left[q\left(\frac{1}{T}-\frac{1}{T_0}\right)\right]\,.
\label{eq04}
\end{equation}

The instantaneous flux of molar fraction $X$ in bulk obeys
\begin{equation}
\vec{J}=-DX\left(
\frac{\nabla X}{X}+\alpha\frac{\nabla T}{T}\right)\,,
\label{eq05}
\end{equation}
where $\alpha$ is the thermodiffusion
constant~\cite{Bird-Stewart-Lightfoot-2007}. The importance of
thermal diffusion was demonstrated for
gases~\cite{Goldobin-Brilliantov-2011} and methane
hydrate~\cite{Goldobin-CRM-2013,Goldobin-etal-EPJE-2014} on
geological time scales.

Generally, not only solubility but all material properties of the
system depend on temperature and pressure. However, feasible
amplitudes of relative variations of the absolute temperature are
below $5\%$. Hence, one can neglect variation of those
characteristics which depend on temperature polynomially and
consider variation of only those characteristics which depend on
temperature exponentially: solubility (\ref{eq03}) and the
molecular diffusion coefficient $D$. Further, the only
characteristic sensitive to pressure below hundreds of atmospheres
is gas solubility.

\begin{table}[t]
\caption{Chemical physical properties of nitrogen, oxygen, methane,
and carbon dioxide (see text for details). Eq.\,(\ref{eq03}) with
$q$ and $X^{(0)}(T_0,P_0)$ specified in the table matches the
experimental data
from~\cite{solubility-1,solubility-2,solubility-3}.
Eq.\,(\ref{eq06}) with $R_d$ and $\nu$ specified matches the
experimental data
from~\cite{diffusion-1,diffusion-2,diffusion-3}.}
\begin{center}
\begin{tabular}{cp{0.7cm}p{0.7cm}p{0.7cm}p{0.55cm}}
\hline\hline
 & $\mathrm{N_2}$ & $\mathrm{O_2}$ & $\mathrm{CH_4}$ & $\mathrm{CO_2}$
 \\
\hline
$q=-G_i/k_\mathrm{B}$ (K)
 & 781 & 831 & 1138 & 1850 \\[5pt]
$X^{(0)}(20^\circ\mathrm{C},1\,\mathrm{atm})$ ($10^{-5}$)
 & 1.20 & 2.41 & 2.60 & 68.7 \\[5pt]
$R_d$ ($10^{-10}\,\mathrm{m}$)
 & 1.48 & 1.29 & 1.91 & 1.57 \\[5pt]
$\nu$ ($10^{-5}\,\mathrm{Pa\cdot s}$)
 & 9.79 & 16.3 & 28.3 & 4.68 \\[3pt]
\hline\hline
\end{tabular}
\end{center}
\label{params}
\end{table}

To complete the mathematical formulation of the problem, one needs
to specify the dependence of molecular diffusion on
temperature~\cite{Bird-Stewart-Lightfoot-2007}
\begin{equation}
D=\frac{k_\mathrm{B}T}{2\pi\mu R_d}\cdot
 \frac{\mu+\nu}{2\mu+3\nu}\,,
\label{eq06}
\end{equation}
where $k_\mathrm{B}$ is the Boltzmann constant, $\mu$ is the dynamic viscosity of the solvent, $R_d$ is the effective radius of the solute molecules with the ``coefficient of sliding friction'' $\beta$, $\nu=R_d\beta/3$. Eq.\,(\ref{eq06}) describes the bulk molecular diffusion without contributions from the effects which become important when the molecule free path is commensurable to the pore size (e.g., \cite{Berson-Choi-Pharoah-2011}). The dependence of dynamic viscosity on temperature can be described by a modified Frenkel formula~\cite{Frenkel-1955}
\begin{equation}
\mu=\mu_0\exp\frac{a}{T+\tau}\,.
\label{eq07}
\end{equation}
For water, coefficient
 $\mu_0=2.42\cdot10^{-5}\,\mathrm{Pa\cdot s}$,
 $a=W/k_\mathrm{B}=570\,\mathrm{K}$
($W$ is activation energy), and $\tau=-140\,\mathrm{K}$.

\begin{figure}[!t]
\center{
\includegraphics[width=0.425\textwidth]%
 {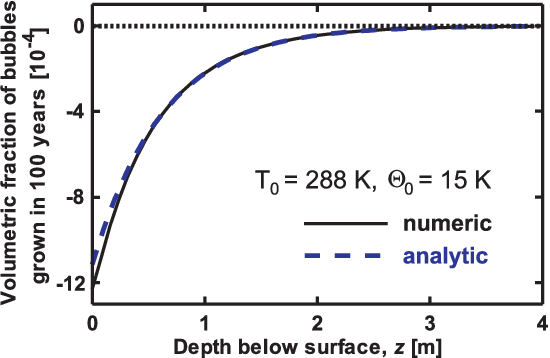} }

  \caption{(Color online)
Comparison of the numerically evaluated divergence of mean flux
with diffusion coefficient (\ref{eq06}) and analytical expressions
(\ref{eq10}) and (\ref{eq11}) for annual temperature wave.
Negative values of the growth rate in this figure mean dissolution
of the bubbly phase.
 }
  \label{fig2}
\end{figure}

\begin{figure*}[!t]
\center{
\begin{tabular}{cc}
{\sf (a)}\hspace{-15pt}
\includegraphics[width=0.435\textwidth]
 {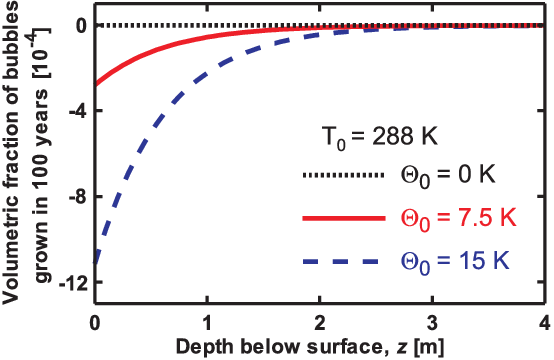}
\qquad
 &
\qquad
{\sf (b)}\hspace{-15pt}
\includegraphics[width=0.435\textwidth]
 {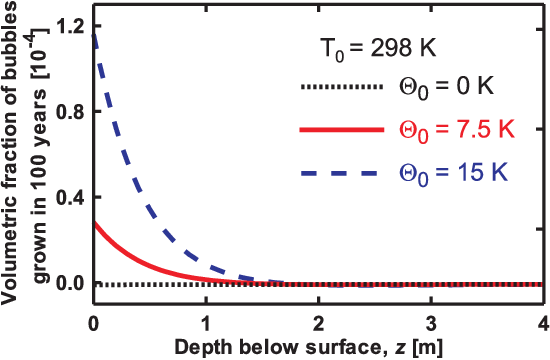}
\\[15pt]
{\sf (c)}\hspace{-15pt}
\includegraphics[width=0.435\textwidth]
 {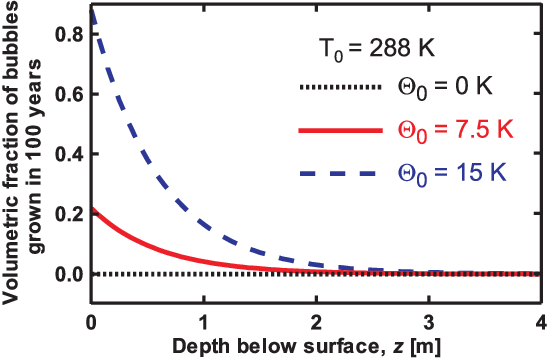}
\qquad
 &
\qquad
{\sf (d)}\hspace{-15pt}
\includegraphics[width=0.435\textwidth]
 {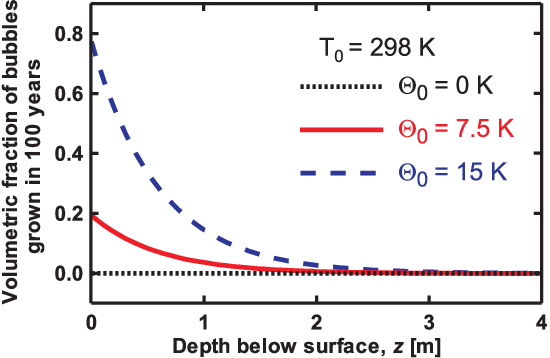}
\end{tabular}}

  \caption{(Color online)
The rate of bubble growth from saturated aqueous solution of
methane subject to annual temperature wave. The transport
mechanism is the molecular diffusion under annual-mean temperature
$T_0=288\,\mathrm{K}$ (a) and $T_0=298\,\mathrm{K}$ (b), and the
hydrodynamic dispersion under annual-mean temperature
$T_0=288\,\mathrm{K}$ (c) and $T_0=298\,\mathrm{K}$ (d). The
surface temperature oscillation amplitude $\Theta_0$ is specified
in plots. For molecular diffusion, rough theoretical assessment
$\alpha=1.8$~\cite{Goldobin-Brilliantov-2011} is adopted, although
parametric study reveals the result to be nearly insensitive to
the specific realistic value of $\alpha$. For hydrodynamic
dispersion $D_\mathrm{vert}=2\cdot10^{-7}\,\mathrm{m^2/s}$, which
typically corresponds, e.g., to the horizontal filtration flux of
groundwater $v_\mathrm{aq}=0.02\,\mathrm{cm/s}$ through the medium
with the pore size $\sim1\,\mathrm{mm}$
(cf.\ \cite{Donaldson-etal-1997,Donaldson-etal-1998}).
 }
  \label{fig3}
\end{figure*}

\section{Average gas mass transport}
Since the diffusion transport in liquids is several orders of
magnitude slower than the heat transfer, it can be well described
in terms of average values over the temperature oscillation
period. With the solubility law (\ref{eq03}) diffusion flux
(\ref{eq05}) reads
\begin{equation}
\vec{J}=-DX^{(0)}\left[
 \frac{\nabla P}{P}-\left(1+\frac{q}{T}-\alpha\right)
 \frac{\nabla T}{T}\right]\,.
\label{eq08}
\end{equation}
With expansion
\begin{equation}
D=D_0+D_1(T-T_0)+\frac{1}{2}D_2(T-T_0)^2+\dots\,,
\label{eq09}
\end{equation}
where $D_0$, $D_1$, and $D_2$ can be plainly evaluated from
eqs.\,(\ref{eq06}) and (\ref{eq07}), straightforward but laborious
analytical calculations (see Appendix for details) yield mean flux
\begin{align}
\left\langle J\right\rangle
 =&-D_0X^{(0)}(T_0,P_0)\bigg[
 \frac{\rho g}{P_0}\left(1+A_1\frac{\Theta_0^2\,e^{-2kz}}{2T_0^2}\right)
 -\left(1+\frac{\rho gz}{P_0}\right)A_2\frac{k\Theta_0^2\,e^{-2kz}}{2T_0^2}
 +\mathcal{O}\left(\frac{\Theta_0^4}{T_0^4}\right)
 \bigg]\,,
 \label{eq10}
\end{align}
where
$$
A_1=1+2\frac{q}{T_0}+\frac{q^2}{2T_0^2}
 -\frac{D_1T_0}{D_0}\left(1+\frac{q}{T_0}\right)
 +\frac{D_2T_0^2}{2D_0}\,,
$$
$$
A_2=\frac{q}{T_0}+\left(2+\frac{q}{T_0}-\frac{D_1T_0}{D_0}\right)
 \left(1+\frac{q}{T_0}-\alpha\right).
$$
We do not provide analytical expressions for $D_1$ and $D_2$ as they are extremely lengthy. These values strongly depend on temperature; nonetheless, to see the reference values of these parameters, one can evaluate from eqs.\,(\ref{eq06}) and (\ref{eq07}) for $T_0=300\,\mathrm{K}$ and data in table~\ref{params}: $D_1T_0/D_0=7.39$, $7.26$, $7.12$, $7.52$ and $D_2T_0^2/D_0=28.1$, $26.6$, $25.2$, $30.2$ for $\mathrm{N_2}$, $\mathrm{O_2}$, $\mathrm{CH_4}$, $\mathrm{CO_2}$, respectively.

Beyond the penetration zone of the decaying temperature wave,
i.e., for $kz\gtrsim2$, the diffusion flux (\ref{eq10}) is
homogeneous, no gas mass accumulation or substraction is created
by the flux divergence. Meantime, in the near-surface zone where
the temperature wave is non-small, the flux divergence is non-zero
and one observes the growth (or dissolution) of the gaseous phase;
\begin{equation}
\frac{\partial\left\langle X_b\right\rangle}{\partial t}=
 -\frac{\partial\left\langle{J}\right\rangle}{\partial z}\,.
\label{eq11}
\end{equation}
Here $X_b$ is the molar amount of matter in the gaseous phase per
1 mole of all the matter in pores. The latter equation is valid
for small volumetric fraction of bubbles in pores, which well
corresponds to the systems we consider.

In fig.\,\ref{fig2}, one can see that analytical results involving the expansion (\ref{eq09}) match well the results of numerical calculations, in which the expressions for the diffusion coefficient and solubility are not replaced by their truncated Taylor series, even for surface temperature oscillation amplitude $\Theta_0$ as large as $15\,\mathrm{K}$. Henceforth, our treatment relies entirely on these analytical results.

\section{Molecular diffusion and hydrodynamic dispersion}
Up to this point our results are derived for the molecular
diffusion mechanism. With molecular diffusion, terms
$(D_1T_0/D_0)$ and $(D_2T_0^2/D_0)$ are of the order of magnitude
of $10$~\footnote{For instance, $(D_1T_0/D_0)$ and
$(D_2T_0^2/D_0)$ are within the range $20$--$30$ for methane as
can be calculated with eqs.\,(\ref{eq03}), (\ref{eq06}),
(\ref{eq07}) and data from table~\ref{params}.}; they make
principle contribution to constants $A_1$ and $A_2$, and play a
decisive role in the systems evolution. Noticeably, calculations
reveal the effect of thermodiffusion ($\alpha$) to be practically
unobservable against the background of other contributions to the
diffusion flux. Hence, the uncertainty of the value of $\alpha$
for aqueous gas solutions~\footnote{Up to authors' knowledge,
neither experimental data nor reliable theoretical calculation of
thermodiffusion constant $\alpha$ for weakly solvable gases in
liquid water can be found in the literature.} is not a significant
issue for the results provided in this paper.

However, along with the molecular-diffusion dominated systems,
there are systems where horizontal filtration flux of groundwater
is present. This flux is treated as horizontal because it is
parallel to the surface and natural systems are typically much
more uniform in the horizontal directions than in the vertical
one. Due to the microscopic irregularity of pore geometry the
filtration flux does fluid mixing which operates like an
additional diffusion mechanism, the ``hydrodynamic
dispersion''~\cite{Donaldson-etal-1997,Donaldson-etal-1998,Sahimi-1993,Barenblatt-Yentov-Ryzhik-2010}.
Although hydrodynamic dispersion was previously studied in the
relation to the evolution of atmosphere gases bubbles in
aquifers~\cite{Donaldson-etal-1997,Donaldson-etal-1998}, the
effect of temperature waves was not addressed.

When hydrodynamic dispersion is present, it is typically
several orders of magnitude stronger than molecular diffusion
(\cite{Donaldson-etal-1997,Donaldson-etal-1998,Barenblatt-Yentov-Ryzhik-2010},
cf.\ in fig.\,\ref{fig3}), and the latter can be neglected. There
is no analog of thermodiffusion for hydrodynamic dispersion and
$\alpha=0$ for this case. Furthermore, hydrodynamic dispersion
depends on the flux strength and pore geometry but not
temperature, and the correlations between instantaneous flux
oscillations and temperature are not obvious; therefore, $D=D_0$
and $D_1=D_2=0$. Finally, for {\em hydrodynamic-dispersion
dominated system}, eqs.\,(\ref{eq10}) and (\ref{eq11}) yield
\begin{align}
\frac{\partial\left\langle X_b\right\rangle}{\partial t}=
 & D_0X^{(0)}(T_0,P_0)\left(1+\frac{2q}{T_0}+\frac{q^2}{2T_0^2}\right)
 \frac{\Theta_0^2}{2T_0^2}
 \frac{\partial^2}{\partial z^2}
 \left[\left(1+\frac{\rho gz}{P_0}\right)e^{-2kz}\right]
 +\mathcal{O}\left(\frac{\Theta_0^4}{T_0^4}\right).
\label{eq12}
\end{align}

In this paper we analyze both molecular-diffusion dominated
systems and hydrodynamic-dispersion dominated systems.

\section{Results and discussion}
In fig.\,\ref{fig3} one can see the results of calculations of the
methane bubble growth rate in molecular-diffusion and
hydrodynamic-dispersion dominated systems. These plots demonstrate
several unique features of these cases. Let us discuss these
features and their origins in detail.

Noteworthy, with molecular diffusion of methane, one can observe
both bubble dissolution for lower mean temperatures
(fig.\,\ref{fig3}a) and bubble growth for higher mean temperatures
(fig.\,\ref{fig3}b). This is dominantly controlled by parameters
$(D_1T_0/D_0)$ and $(D_2T_0^2/D_0)$ and may vary from gas to gas.
Analytical expression (\ref{eq10}) contains both exponential and
linear in $z$ terms and multipliers. With purely exponential
terms, the shape of the bubble growth rate profile would be
determined only by $k$ [see eq.\,(\ref{eq10})]. Owing to the
linear in $z$ multiplier in eq.\,(\ref{eq10}), this tolerance of
the profile shape to parameters $A_1$ and $A_2$ is broken by the
competition between terms $e^{-2kz}$ and $ze^{-2kz}$. In
particular, figs.\,\ref{fig3}a and~\ref{fig3}b demonstrate that
the penetration depth of the bubble depletion zone is nearly twice
bigger than the penetration depth of the growth zone.

On the contrary, for hydrodynamic-dispersion dominated systems,
only growth of bubbles is possible. According to
eq.\,(\ref{eq12}), the shape of the bubble growth rate profile is
determined by the second $z$-derivative term. This derivative
profile depends on $k=\sqrt{\omega/2\chi}$. The second
$z$-derivative is everywhere positive for $k>\rho
g/P_0=0.1\,\mathrm{m^{-1}}$, i.e., for temperature wave
penetration depth below $10\,\mathrm{m}$, while for $k<\rho g/P_0$
it turns negative in the upper zone above $z_*=k^{-1}-P_0/\rho g$.
Since for natural systems the penetration depth of the annual
temperature wave never reaches
$10\,\mathrm{m}$~\cite{Yershov-1998}, the latter case, with bubble
depletion above $z_*$ and growth below $z_*$, is not feasible and
can be relevant only for longer climate cycles.

We discussed our analytical results in relation to geological
systems featured by the hydrostatic pressure gradient and annual
temperature oscillations. Meanwhile, eqs.\,(\ref{eq10}) and
(\ref{eq12}) hold valid for the systems where the hydrostatic
pressure gradient is either absent or insignificant. The latter
case may be, for example, daily temperature oscillation, the
penetration depth of which is $0.1\,\mathrm{m}$ for soils---a
depth on which the hydrostatic pressure increase is negligible.
The case of {\em no pressure gradient} can be derived from
eq.\,(\ref{eq10}) by eliminating the gravity; one finds
\begin{equation}
\left\langle J\right\rangle
 =D_0X^{(0)}(T_0,P_0)\,A_2\frac{k\Theta_0^2\,e^{-2kz}}{2T_0^2}
 +\mathcal{O}\left(\frac{\Theta_0^4}{T_0^4}\right)\,.
 \label{eq13}
\end{equation}
Here we can see a non-trivial temperature wave effect on gas
release for various technological and natural systems as well.

For solids, the results correspond to the case of no pressure
gradient, eq.\,(\ref{eq13}), with correction of the coefficient
$A_2$ due to the solubility law (\ref{eq04}), which differs
from eq.\,(\ref{eq03}) for gases;
\begin{equation}
A_{2,\mathrm{solid}}=\frac{q}{T_0}+\left(1+\frac{q}{T_0}-\frac{D_1T_0}{D_0}\right)
 \left(1+\frac{q}{T_0}-\alpha\right).
 \label{eq14}
\end{equation}
Thus, one can observe or achieve the same effect of a spatially
distributed growth of the non-dissolved phase in porous matrix.

\subsection*{Controlling mass distribution pattern}
One can control the pattern of the growth of the non-dissolved phase. Let us discuss this possibility for the case of solid substances.
According to eq.\,(\ref{eq13}), the surface temperature
oscillation waveform
$$
\Theta(z=0,t)=\int_{-\infty}^{+\infty}
 \theta(\omega)\,e^{i\omega t}\mathrm{d}\omega
$$
produces the average growth rate
\begin{align}
 -\frac{\partial\langle{J}\rangle}{\partial z}&=
 \frac{4D_0X^{(0)}(T_0,P_0)\,A_2}{T_0^2}\int_0^{+\infty}
 k^2|\theta(\omega)|^2\,e^{-2kz}\mathrm{d}\omega
 \nonumber\\[5pt]
 &=
 \frac{\chi D_0X^{(0)}(T_0,P_0)\,A_2}{T_0^2}\int_0^{+\infty}
 \kappa^3\left|\theta\left(\frac{\chi\kappa^2}{2}\right)\right|^2\,e^{-\kappa z}\mathrm{d}\kappa\,.
 \label{eq15}
\end{align}
(The cross-terms of different frequencies, $\propto\theta(\omega)\theta(-\omega')$ with $\omega\ne\omega'$, yield vanishing time-average contributions.) The integral in the latter expression has a form of a Laplace integral, which can be used for representation of a function defined for $z\ge0$, $f(z)=\int_0^{+\infty}F(\kappa)e^{-\kappa z}\mathrm{d}\kappa$ with real positive $F(\kappa)$; here $F(\kappa)$ is the inverse Laplace transform of function $f(z)$. Let us first consider the case of $A_2>0$. If one requires a growth pattern the inverse Laplace transform $F(\kappa)$ of which is real positive, i.e.,
\begin{align}
-\frac{\partial\langle{J}\rangle}{\partial z}\propto
 \int_0^{\infty}a(\kappa)\,e^{-\kappa z}\,\mathrm{d}\kappa
 \label{eq16}
\end{align}
with $a(\kappa)\ge 0$, this can be achieved by implementation of
the surface temperature oscillation waveform
 $\Theta(0,t)=\int_{-\infty}^{+\infty}
 \theta(\omega)e^{i\omega t}\mathrm{d}\omega$,
producing pattern (\ref{eq15}). Comparing integrals in eqs.\,(\ref{eq15}) and (\ref{eq16}), one finds $a(\kappa)\propto\kappa^3\left|\theta\left(\chi\kappa^2/2\right)\right|^2$, which corresponds to
\begin{align}
 \theta(\omega)\propto\omega^{-3/4}\big[a\big(\sqrt{2\omega/\chi}\big)\big]^{1/2}\,.
 \label{eq17}
\end{align}
For $A_2<0$, similar depletion pattern, with $(-\partial\langle{J}\rangle/\partial z)<0$, can be created.
Thus, a certain class of patterns can be produced by means of a
proper temperature oscillation waveform. Presumably, this can be
generalized to a spatially non-unform surface temperature
oscillations, allowing producing a class of three-dimensional
patterns of the ``guest'' substance in porous matrix.

\section{Conclusion}
We have addressed the problem of influence of the surface
temperature oscillation (and consequent temperature wave) on the
gas transport through liquid-saturated porous media. Specifically,
we have considered the problem for the case of saturated gas
solution, i.e., when bubbles are spread everywhere in the medium.
Strong exponential dependence of the solubility and the molecular
diffusion coefficient on temperature has been found to lead to
non-negligible instantaneous diffusion fluxes although the
amplitude of the relative variation of absolute temperature does
not exceed $5\%$. Due to nonlinearity these fluxes are not
averaged-out to zero but create the mean mass flux.

Our main findings are expressed by eq.\,(\ref{eq10}) and its
reductions for hydrodynamic-dispersion dominated systems
[eq.\,(\ref{eq12})] and no pressure gradient systems
[eq.\,(\ref{eq13})]. These analytical equations have been shown to
fairly match the results of numerical calculations (see
fig.\,\ref{fig2}). We have revealed that the temperature wave can
either enhance or deplete the near-surface bubbly zone in
molecular-diffusion dominated systems and only enhance this zone
for hydrodynamic-dispersion dominated systems (for instance, see
fig.\,\ref{fig3}).

The phenomenon we have addressed is expected to be of significance
not only for gases but also for any weakly-soluble substance,
given its solubility is sensitive to temperature. Moreover, one
can control the pattern of the growth of the non-dissolved phase.

\begin{acknowledgement}
Authors acknowledge financial support by the Government of Perm
Region (Contract C-26/212) and the Russian Foundation for Basic
Research (project no. 14-01-31380\_mol\_a).
\end{acknowledgement}

\appendix
\section*{Appendix: Calculation of average diffusion flux}
Here we derive eq.\,(\ref{eq10}) for the average diffusion flux from eq.\,(\ref{eq08}) using expression for the solubility~(\ref{eq03}) and diffusion coefficient~(\ref{eq09});
\begin{eqnarray}
\langle J\rangle&=&
 -\bigg\langle
  \left(D_0+D_1(T-T_0)+\frac{1}{2}D_2(T-T_0)^2+\dots\right)
\nonumber\\[5pt]
&&\qquad\times
 X^{(0)}(T_0,P_0)\frac{T_0}{T}\frac{P}{P_0}
 \exp\left[q\left(\frac{1}{T}-\frac{1}{T_0}\right)\right]
 \left[\frac{\nabla P}{P}-\left(1+\frac{q}{T}-\alpha\right)
 \frac{\nabla T}{T}\right]\bigg\rangle\,.
\label{eqapp1}
\end{eqnarray}
With short-hand notation $T=T_0(1+\vartheta)$, making expansions in $\vartheta$, one finds
\begin{eqnarray}
 \frac{\langle J\rangle}{D_0X^{(0)}(T_0,P_0)}&=&
 -\bigg\langle
  \left(1+\frac{D_1T_0}{D_0}\vartheta+\frac{D_2T_0^2}{2D_0}\vartheta^2\right)
  \left(1-\frac{q}{T_0}\vartheta+\left(\frac{q}{T_0}
 +\frac{q^2}{2T_0^2}\right)\vartheta^2\right)
\nonumber\\[5pt]
&&\qquad\times
 \bigg(\frac{\rho g}{P_0}(1-\vartheta+\vartheta^2)
 -\frac{P}{P_0}\left[(1-\alpha)(1-2\vartheta)+\frac{q}{T_0}(1-3\vartheta)\right]
 \nabla\vartheta\bigg)
 +\mathcal{O}(\vartheta^3)\bigg\rangle\,.
\label{eqapp2}
\end{eqnarray}
As $\langle\vartheta\rangle=0$, $\langle\vartheta^2\rangle=\frac{\Theta_0^2}{2T_0^2}e^{-2kz}\ne0$, $\langle\vartheta\nabla\vartheta\rangle=-\frac{k\Theta_0^2}{2T_0^2}e^{-2kz}\ne0$, and $\langle\vartheta^3\rangle=0$, eq.\,(\ref{eqapp2}) yields
\begin{eqnarray}
 \frac{\langle J\rangle}{D_0X^{(0)}(T_0,P_0)}&=&
 -\frac{\rho g}{P_0}
 -\frac{\rho g}{P_0}\langle\vartheta^2\rangle
 \left(1+2\frac{q}{T_0}+\frac{q^2}{2T_0^2}-\frac{D_1T_0}{D_0}-\frac{D_1q}{D_0}
 +\frac{D_2T_0^2}{2D_0}\right)
\nonumber\\[5pt]
&&\qquad
 {}-\frac{P}{P_0}\langle\vartheta\nabla\vartheta\rangle
 \bigg[2-2\alpha+3\frac{q}{T_0}
 +\left(\frac{q}{T_0}-\frac{D_1T_0}{D_0}\right)
 \left(1-\alpha+\frac{q}{T_0}\right)\bigg]
 +\mathcal{O}(\langle\vartheta^4\rangle)
\nonumber\\[5pt]
&=&-\frac{\rho g}{P_0}\big(1+A_1\langle\vartheta^2\rangle\big)
 -\frac{P}{P_0}A_2\langle\vartheta\nabla\vartheta\rangle
 +\mathcal{O}(\langle\vartheta^4\rangle)\,.
\nonumber
\end{eqnarray}
with $A_1$ and $A_2$ specified below eq.\,(\ref{eq10}).

For solids, with solubility (\ref{eq04}), one similarly finds
\begin{eqnarray}
 \frac{\langle J\rangle}{D_0X^{(0)}(T_0,P_0)}&=&
 -\frac{P}{P_0}\langle\vartheta\nabla\vartheta\rangle
 \bigg[1-\alpha+2\frac{q}{T_0}
 +\left(\frac{q}{T_0}-\frac{D_1T_0}{D_0}\right)
 \left(1-\alpha+\frac{q}{T_0}\right)\bigg]
 +\mathcal{O}(\langle\vartheta^4\rangle)
\nonumber\\[5pt]
&=& -A_{2,\mathrm{solid}}\langle\vartheta\nabla\vartheta\rangle
 +\mathcal{O}(\langle\vartheta^4\rangle)\,.
\nonumber
\end{eqnarray}
with $A_{2,\mathrm{solid}}$ specified by eq.\,(\ref{eq14}).

\end{document}